\documentclass{iaus}
\usepackage{graphicx}
\usepackage{eurosym}
\usepackage{psfig}
\usepackage[authoryear,square]{natbib}
\bibpunct{(}{)}{;}{a}{}{,}

\title[Cosmic Magnetism with the SKA and its Pathfinders] 
{Cosmic Magnetism with the Square Kilometre Array and its Pathfinders}

\author[B.\ M.\ Gaensler]  
{Bryan M.\ Gaensler$^1$ }

\affiliation{$^1$Sydney Institute for Astronomy (SIfA),
School of Physics, \\ The University of Sydney, NSW 2006, Australia
\break email: bgaensler@usyd.edu.au}

\pubyear{2009}
\volume{259}  
\pagerange{100--100}
\date{December 16th, 2008}
 \jname{Cosmic Magnetic Fields: From Planets,
to Stars and Galaxies} \editors{K.G. Strassmeier, A.G. Kosovichev \&
J.E. Beckman, eds.}

\def\farcs{\hbox{$.\!\!^{\prime\prime}$}}

\begin{document}

\maketitle

\begin{abstract}
One of the five key science projects for the Square Kilometre Array
(SKA) is ``The Origin and Evolution of Cosmic Magnetism'', in which radio
polarimetry will be used to reveal what cosmic magnets look like and what
role they have played in the evolving Universe. Many of the SKA prototypes
now being built are also targeting magnetic fields and polarimetry as key
science areas. Here I review the prospects for innovative
new polarimetry and Faraday rotation experiments with forthcoming
facilities such as ASKAP, LOFAR, the ATA, the EVLA, and ultimately the SKA.
Sensitive wide-field polarisation surveys with these telescopes will
provide a dramatic new view of magnetic fields in the Milky Way, in
nearby galaxies and clusters, and in the high-redshift Universe.

\end{abstract}

\firstsection 
\section{Introduction}

\begin{figure}
\centerline{\psfig{file=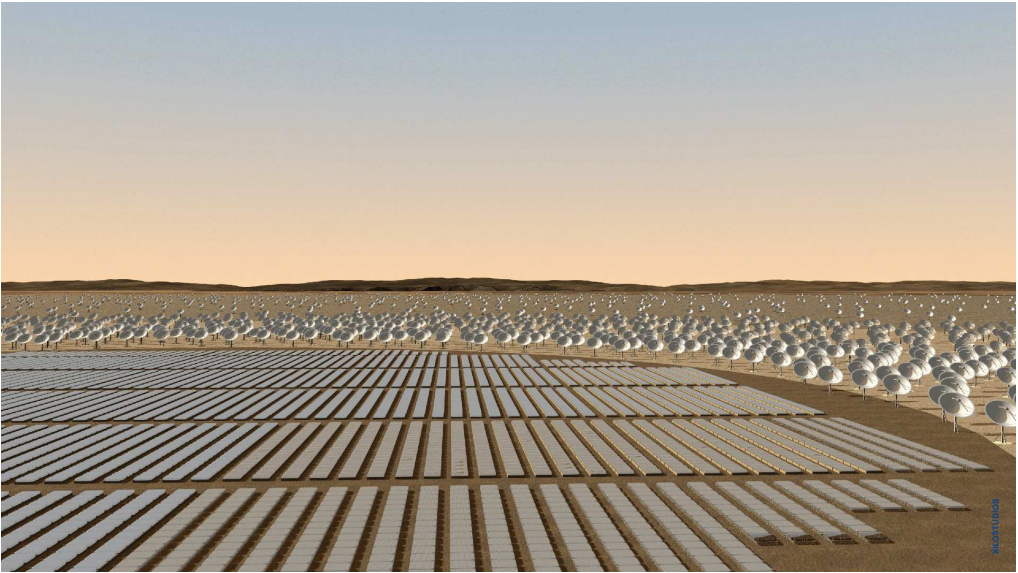,width=\textwidth}}
\caption{An artist's impression of the SKA, showing the central core of
steerable dishes and passive aperture tiles. Image created by XILOSTUDIOS
for the SKA Project Development Office.}
\label{fig_ska}
\end{figure}

The Square Kilometre Array (SKA)\footnote{http://www.skatelescope.org}
is a concept for a next-generation
radio telescope, with a total approximate collecting area of 1~km$^2$
\citep{sdl08}; an artist's impression of the central core
of the SKA is shown in Figure~\ref{fig_ska}.
From the outset, the
SKA project has been a global effort,
and is currently governed by an international consortium of 18
member countries. Two potential sites are being considered 
for the SKA: near Boolardy station in Western Australia, and in
the Karoo region of South Africa. 

The SKA will be a highly flexible instrument, covering a broad range
of frequencies (0.07--25~GHz), with many different operating modes.
At an observing frequency of 1.4~GHz, the SKA will have a maximum
angular resolution of $0\farcs02$ and a field of view of 20~square
degrees. This latter specification provides incredible survey
capability, far exceeding that of any other sensitive radio telescope.
A unique capability of radio arrays is that they can begin taking
science data science long before the full facility is complete.
Operations for the SKA are thus projected to begin in around 2016,
with an approximate construction cost of  \euro 1.5 billion.

Five key science projects have been designated for the SKA, as
discussed by \cite{gae04c} and \cite{cr04}. Below I summarise
recent progress and developments on one of these five key
projects, ``The Origin and Evolution of Cosmic Magnetism''
\citep[see][]{gbf04}.


\section{SKA Polarisation Pathfinders}

Although the full SKA is still some years away, a large number of
pathfinder facilities are either
under construction or are already beginning to take data.
Many of these will be
carrying out exciting new experiments on polarimetry,
Faraday rotation and magnetic fields. New telescopes
with such capabilities include:
\begin{itemize}
\item The Galactic Arecibo L-Band Feed Array Continuum Transit Survey
(GALFACTS),\footnote{http://www.ucalgary.ca/ras/GALFACTS} a 1.4-GHz
survey that began in late-2008, and which will map the entire polarised sky
visible to Arecibo;
\item The Low Frequency Array (LOFAR),\footnote{http://www.lofar.org}
currently under construction in the Netherlands and Germany, which will
study polarisation over the whole northern sky at very low frequencies
($\nu =30-80, 110-240$~MHz) \citep{bec09};
\item The Allen Telescope Array
(ATA)\footnote{http://ral.berkeley.edu/ata} in northern California,
a newly operational facility that has a wide field of view (5~deg$^2$
at 1.4~GHz) and can carry out very large continuum surveys;
\item The Square Kilometre Array Molonglo Prototype
(SKAMP),\footnote{http://www.physics.usyd.edu.au/sifa/Main/SKAMP}
a refurbishment of the Molonglo Observatory Synthesis Telescope in
south-eastern Australia, which will provide 18\,000~m$^2$ of collecting
area for studying diffuse polarisation at
an observing frequency of $\sim1$~GHz over
wide fields;
\item The Murchison Widefield Array
(MWA),\footnote{http://www.haystack.mit.edu/ast/arrays/MWA} an interferometer
being built in Western Australia, which will study polarised
emission over wide fields in the frequency range 80--300~MHz;
\item The Expanded Very Large Array
(EVLA),\footnote{http://www.aoc.nrao.edu/evla}, a substantial upgrade to
the VLA in New Mexico, providing greatly improved continuum sensitivity,
frequency coverage and correlator capability;
\item The Karoo Array Telescope
(MeerKAT),\footnote{http://www.ska.ac.za/meerkat} an array of 80 12-metre
dishes, each equipped with a wideband feed covering the frequency range
0.7--10~GHz;
\item The Australian SKA Pathfinder
(ASKAP)\footnote{http://www.atnf.csiro.au/projects/askap}, an array of 36
12-metre antennas to be built on the Western Australian SKA site. ASKAP
will be a very wide-field survey instrument (30~deg$^2$ at 1.4~GHz),
and will be able to study polarisation at a range of spatial scales
in the frequency range 700--1800~MHz \citep{jbb+07}.

\end{itemize}

\section{Cosmic Magnetism with the SKA}

The SKA key science project on cosmic magnetism focuses on three
themes: structure,  evolution and origin of magnetic fields
\citep{gbf04}.  The questions we hope to address for each theme can
be summarised as follows:
\begin{enumerate}
\item {\bf Structure:} What is the strength and structure of
magnetic fields in the Milky Way, in other galaxies, and in
galaxy clusters?
\item {\bf Evolution:} How have magnetic fields evolved in galaxies
and clusters over cosmic time?
\item {\bf Origin:} When and how was the Universe magnetised?
\end{enumerate}

In the following discussion, we outline experiments with
the SKA and its pathfinders that can address each of these topics.

\subsection{Structure: The Rotation Measure Grid}
\label{sec_grid}

Recent surveys of polarised extragalactic sources, carried out with the
Australia Telescope Compact Array and the DRAO Synthesis Telescope, have
yielded background Faraday rotation data at a sky density of $\sim1$~RM
per deg$^2$.  These studies have started to reveal the large-scale
magnetic field geometry of the Milky Way and Magellanic Clouds, allowing
us to infer parameters such as the pitch angle of the magnetic field,
the presence and location of field reversals, and the overall dynamo
mode \citep[e.g.,][]{ghs+05,bhg+07,mgs+08}.

Observations with the SKA can spectacularly improve the sky density
of background RMs. An hour of integration with the full SKA
will yield approximately 2000 RMs per deg$^2$.  Over the full 20~deg$^2$
field of view, we will obtain an order of magnitude more background RMs
in the first hour of observations than what has been accumulated over
the last 40 years!

\begin{figure}
\centerline{\psfig{file=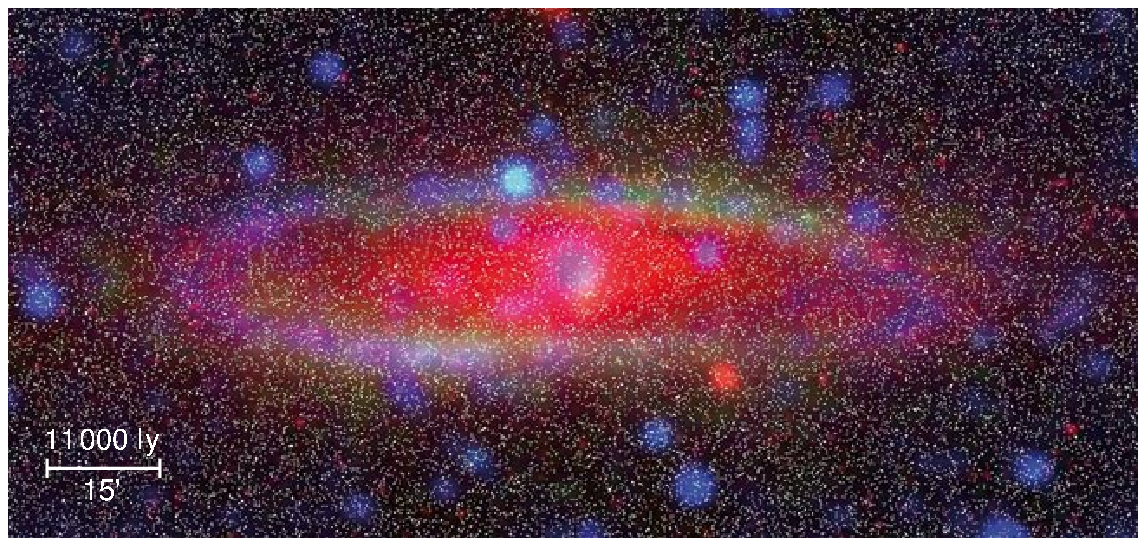,width=\textwidth,clip=}}
\caption{A simulation of the rotation measure grid behind the nearby
galaxy M31, as seen with the SKA in a 30-hour observation.  The colours
are overlays from three observations: red indicates optical emission,
blue corresponds to total intensity radio emission
at 5~GHz, and green indicates polarised radio emission at 5~GHz.
The white dots show the locations of 50\,000 
simulated polarised background sources
for which the SKA could extract RMs.
Optical image: Digitized Sky Survey; radio images: courtesy of Rainer Beck.}
\label{fig_m31}
\end{figure}

This will allow many exciting new applications of the ``rotation
measure grid''.  For our own Milky Way, we will be able to combine
these data with pulsar RMs to derive a full three-dimensional map
of the Galactic magnetic field, on scales ranging from the overall
geometry of the field in the disk and halo, down to the properties
of magnetised turbulence on sub-parsec scales. Deep observations
of nearby galaxies will yield $>10^5$ RMs per target, as shown in
Figure~\ref{fig_m31} --- these data will allow a full reconstruction
of the large-scale magnetic field in these systems. An all-sky
survey of polarised continuum emission with the SKA will provide
$\sim10$ background RMs for each of the $\sim5000$ nearest galaxies,
for each of which we will be able to fit for the overall geometry
and structure of the magnetic field \citep[see][for a full
discussion]{sab+08}.  These studies will deliver a definitive census
on the magnetic field properties of typical galaxies, and will allow
us to explore how field properties (e.g., field strength, pitch
angle, number of reversals) depend on parameters such as galaxy
type, presence/absence of a bar, or degree of interaction.

In the lead-up to the SKA, pathfinder instruments also have an important
role to play. In particular, the new wide-field sensitive surveys
that will be carried out by ASKAP and the EVLA will allow us to derive
catalogues of polarised extragalactic source counts down to fluxes of
a microjansky or lower. The polarisation properties of these sources
will allow us to separate starburst populations from low-luminosity AGNs
\citep{tsg+07}, and to characterise the statistical properties of magnetic
field geometries in unresolved spiral galaxies \citep{abks09,skbt09}.

\subsection{Evolution: Rotation Measure vs.\ Redshift}

For Faraday rotation at high redshifts, RMs in the observer frame will
be reduced in magnitude by a factor $(1+z)^2$ compared to the RM in the
region where rotation occurs. Thus if extragalactic RMs are dominated by
Faraday rotation intrinsic to the emitting source, and if all sources
have similar intrinsic RMs, then we expect that an ensemble of sources
for which we have measured both RM and $z$ will demonstrate a dependence
$|{\rm RM}| \propto (1+z)^{-2}$. However, a sample of 268 high-latitude
quasars in the redshift range $0 < z < 3.7$ show a slight {\em increase}\
in RM as a function of $z$ \citep{kbm+08}. This result, along with other
new measurements of Faraday rotation \citep{bml+08} and Zeeman splitting
\citep{wjr+08} against sources at $z\sim0.5-2$, provides strong evidence
that the observed RMs have substantial contributions from intervening
absorbing galaxies, and that throughout the last 10 Gyr of cosmic time,
typical field strengths in these absorbers have been at the
level of a few microgauss or more. This
is at odds with the expectation from mean-field dynamo theory that
the galactic magnetic fields observed today have steadily grown over
time, and favours models in which the field undergoes relatively rapid
amplification.

Current instrumental capabilities introduce three limitations on these
intriguing new results. First, since the observed RM from an extragalactic
source is usually dominated by the contribution from the Milky Way's
interstellar medium (ISM), any correlation between RM and $z$ must use
not the total observed RM, but the {\em residual}\ rotation measure
(RRM), i.e., the RM after  a contribution from the Galactic foreground
has been estimated and subtracted. Currently, the relatively spare
sampling of RMs over the sky limits the accuracy with which we can
calculate RRMs \citep{dc05,shk07}. The greatly improved RM grid data that
will come from pathfinder surveys with ASKAP and the ATA will provide
a dramatic improvement in our ability to model and remove the Galactic
foreground. The corresponding RRMs will have much smaller uncertainties.

Second, we simply lack the sample sizes needed to build up sufficient
statistics of RM vs.\ $z$. An all-sky polarisation survey with the SKA
down to sub-$\mu$Jy sensitivities will yield $\approx10^8$ background
RMs. Provided that a few percent of  these sources also have spectroscopic
or photometric redshifts, we will be able to invert the distribution of
$({\rm RM}, {\rm Dec.}, {\rm RM}, z)$ to derive the magnetic power spectrum
of the IGM as a function of time, out to $z\sim3$ \citep{kol98,bbo99}.

Finally, if indeed the evolution of magnetic field strengths in
galaxies is relatively slow over the last 10~Gyr, then we need RM
data on very distant sources. Such measurements can allow us to
push back to early enough times to reveal how magnetic fields in
galaxies were created and amplified. Deep SKA observations of quasars
or gamma-ray burst afterglows at $z>6$ can provide information on
Faraday rotation at these early epochs.

\subsection{Origin: The Cosmic Web}

Deep observations of galaxy clusters at low frequencies have begun to
reveal filaments and diffuse synchrotron emission seen on the periphery
of, or even between, clusters \citep{kksp07,pdfg08}.  While the
detections so far have been at relatively low signal-to-noise,
such studies have enormous potential, because they serve
as direct probes of relativistic particles and magnetic fields in the
pristine intergalactic medium (IGM) \citep{hb07,rkcd08}.  Furthermore,
\cite{xkhd06} have suggested that the magnetic field of the IGM might be
measurable through the Faraday rotation of the background RM grid,
as discussed in \S\ref{sec_grid} above.

Observations of total intensity, polarisation and Faraday rotation
with the MWA, LOFAR and SKA can extend such studies far beyond their
current preliminary levels, and will provide superb insights
into the magnetised large-scale structure of the Universe
\citep{kwl04a,kwl04b,bps+09}. In particular, the structure
of the magnetised IGM may serve as a potential discriminant between
primordial models vs.\ outflows as the origin of the IGM magnetic 
field \citep{ddlm09}.

\subsection{Multi-wavelength Synergies}

All the above experiments will have important overlap with new experiments
being carried out in other wavebands and using other techniques. At
much higher radio frequencies than will be observed by the SKA, the
{\em Planck}\ satellite will make a superb Faraday-free map of Galactic
synchrotron emission and its polarisation \citep{ewvs06}. At optical and
infrared wavelengths, forthcoming wide-field survey facilities such as
SkyMapper, PanSTARRS and the LSST will provide complementary information
on large-scale structure and on photometric redshifts. And finally,
deflection of the trajectories of ultra-high-energy cosmic rays measured
by Auger will help map the strength and structure of magnetic fields in
the local IGM.

\section{Conclusions}

There have been many new discoveries and ideas relating to cosmic
magnetism in the last few years, Using these new results as a starting
point, the SKA and its pathfinders promise a suite of unique experiments
aimed at revealing the structure, evolution and origin of magnetic
fields in the Universe.  Specifically, the dense RM grid that the SKA
will provide will probe the structure of magnetic fields in the Milky
Way, nearby galaxies and clusters; studies of RM vs.\ redshift will
allow us to measure the evolution of magnetic fields in galaxies and in
the IGM over cosmic time; and radio imaging of the relativistic cosmic
web provides a direct view of intergalactic magnetic fields, providing
information on the origin of magnetism in the Universe.

The richness of the polarised sky is already beginning to be revealed
by pathfinder experiments such as GALFACTS and the ATA, with even more
powerful facilities such as the EVLA and ASKAP now under construction.
These activities will culminate in the next decade with the
arrival of the SKA, and a consequent exploration
of the full magnetic Universe.  This endeavour will address
major unanswered issues in fundamental physics and astrophysics, and
will almost certainly yield new and unanticipated results.

\begin{acknowledgments}
I thank the organisers for financial support, and my many colleagues
with whom I have collaborated on polarimetry and magnetic fields. I
acknowledge the support of a Federation Fellowship from the Australian
Research Council through grant FF0561298.
\end{acknowledgments}

\bibliographystyle{apj}
\bibliography{journals,modrefs,psrrefs,crossrefs}

\begin{discussion}

\discuss{Kronberg}{The ``Extended Very Large Array (EVLA) II'', whose funding
request was rejected, would serve as an obvious nucleus of a future
``northern SKA'' operating at the higher frequencies.}

\discuss{de Gouveia Dal Pino}{After this excellent talk on SKA and
the coming projects for cosmic magnetism search, I'd like to make
a comment that there is presently a project for building a millimetre
VLBI apparatus in the Southern Hemisphere. It is an Argentinian-Brazilian
collaboration for installing two antennas in desert areas in Argentina
at distances of $\sim100$~km from ALMA, allowing VLBI with ALMA at
mm frequencies.}

\end{discussion}

\end{document}